\documentclass[final,english]{bullsrsl}[2022/06/15]

\usepackage[latin1]{inputenc}
\usepackage[T1]{fontenc}

\usepackage{natbib} 
\usepackage{graphicx}

\begin{document}
\title{Peculiar abundances of the cool carbon star HE 1104-0957}

\author[, corresponding]{Meenakshi}{Purandardas}
\author[]{Aruna}{Goswami}
\author[]{Sridharan}{Rengasamy}
\affiliation[]{Indian Institute of Astrophysics, Koramangala, II block, Bangalore, Karnataka, 560034.}
\correspondance{meenakshi.p@iiap.res.in}
\maketitle
\begin{abstract}
 
We present, for the first time, a detailed abundance analysis of the carbon star HE 1104$-$0957 based on high resolution (R${\sim}$ 50\,000) spectra. Our analysis shows that the object is an extremely metal-poor star with [Fe/H] $\sim$  $-$2.96. We find that the object shows enhancement of carbon with [C/Fe] $\sim$ 1.82. However, it does not fall into any of the sub-groups of carbon-enhanced metal-poor (CEMP) stars based on the characteristic elemental abundance ratios used for the classification of various CEMP sub-groups. HE 1104$-$0957 is also found to exhibit an enhancement of oxygen and nitrogen with [O/Fe], and [N/Fe] $\sim$ 1.54, and 2.54 respectively. In HE 1104$-$0957, $\alpha$-elements are found to be slightly enhanced with [$\alpha$/Fe] $\sim$ 0.46. Fe-peak elements are also moderately enhanced in HE 1104$-$0957 with a value 0.63 with respect to Fe. Our analysis shows that HE 1104$-$0957 exhibits enhancement of neutron-capture elements, particularly r-process elements. The low-resolution spectra of this object shows the spectral features characteristics of a typical C-R star. However, We find that the surface chemical compositions of this object is contradictory to that expected for a C-R star. It requires a detailed analysis to better understand the abundance anomalies exhibited by this object. 
\end{abstract}

\vspace{0.5cm}
\keywords{ Metal-poor stars---Abundances---chemically peculiar---nucleosynthesis}
%
%
\section{Introduction}

Carbon stars are a peculiar class of objects that show strong absorption bands due to carbon molecules such as C$_{2}$, CH, and CN in their spectra. Due to their surface chemical peculiarities, carbon stars have drawn considerable attention as far as their spectroscopic studies are concerned. Carbon stars are classified into different groups such as CH, CN, C-R, and C-J stars based on their spectral features, such as the strength of the C$_{2}$ bands, CN bands, SiC$_{2}$ bands, the strength of Ca I line at 4226 \AA, prominence of the P-branch head around 4342 \AA, etc. Other than the spectral features, these groups also exhibit differences in their surface chemical compositions, effective temperatures, galactic membership, etc. 

Among these sub-classes of carbon stars, C-R and C-J stars are not much explored. For these two groups, detailed studies based on high-resolution spectra are quite scanty in the literature. \cite{Goswami2005} identified the object HE 1104$-$0957 (RA$_{2000}$\,=\,11h07m19.40s; DEC$_{2000}$\,=\,$-$10d13m15.89s) as a C-R star based on low-resolution spectroscopic analysis. We have conducted a high-resolution spectroscopic follow-up of this object with a primary goal to understand the surface chemical composition and hence to probe its possible progenitor. In this paper, we present the abundance analysis results for HE 1104$-$0957. The paper is organized as follows: Section\,2 presents the details on the acquisition of data. Details of our analysis are presented in Section\,3, and the conclusions are drawn in Section\,4.

\section{Acquisition of Data}

The object is selected from \cite{Goswami2005} where the author has identified HE 1104$-$0957 as a C-R star from a set of spectral features  based on the analysis of its low-resolution spectra (R $\sim$ 1330) obtained from the Himalayan Faint Object Spectrograph Camera (HFOSC) attached to the Himalayan Chandra Telescope (HCT), Indian Astronomical Observatory (IAO), Hanle, India. 

\begin{figure}[t]
\centering
\includegraphics[width=0.5\textwidth]{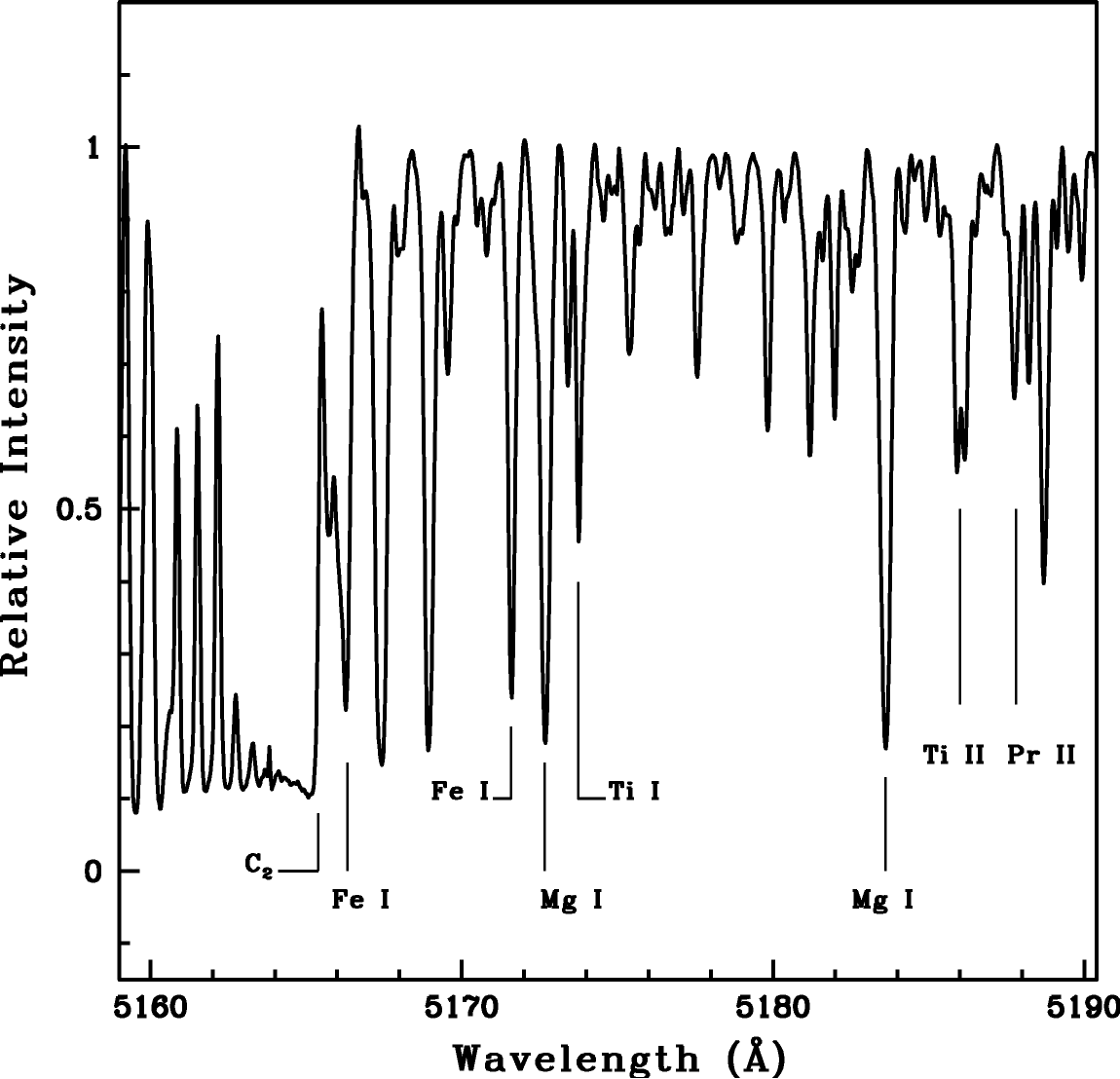}
\bigskip
\begin{minipage}{12cm}
\caption{Sample spectra of HE 1104$-$0957 in the wavelength range 5160 - 5190 \AA}
\label{fig:1}
\end{minipage}
\end{figure}

We obtained the high-resolution spectra (R$\sim$ 50 000) for HE 1104$-$0957 from the Japanese Virtual Observatory (JVO) portal (\url{http://jvo.nao.ac.jp/portal/v2/}) operated by the National Astronomical Observatory of Japan (NAOJ). 
These spectra were acquired using the High Dispersion Spectrograph (HDS) of the 8.2 m Subaru Telescope. The spectra cover a wavelength range that extends from 4100 to 6850{\rm \AA}. There exits a gap between 5440 and 5520{\rm \AA} which arises from the physical separation between the two EEV CCDs with 2048$\times$4096 pixels with two by two on-chip binning. The signal to noise ratio of the spectra of the programme stars range from $\sim$ 53 - 67 per resolution element measured at 6000 \AA. Example of the sample spectra is shown in Fig.\,\ref{fig:1}. The basic data of the program star are presented in Table\,\ref{tab:1}, where the B, V, J, H, and K magnitudes are taken from SIMBAD database.

\begin{table}[t]
\centering
\begin{minipage}{132mm}
\caption{Basic data for HE 0110$-$0947.}
\label{tab:1}
\end{minipage}  
\bigskip

\begin{tabular}{ccccccccc}
\hline
\textbf{B} & \textbf{V} & \textbf{J} & \textbf{H} & \textbf{K} & \textbf{Exposure} & \textbf{S/N}          & \textbf{Date of obs.} \\
           &            &            &            &            & \textbf{(s)}      & \textbf{(at 6000\AA)} &                       \\
\hline
12.12 & 10.76  & 8.26   & 7.56  & 7.31  & 1200 & 65 & 08-12-2003  \\ 
\hline
\end{tabular}
\end{table}

\section{Analysis}

\subsection{Determination of radial velocity}

The radial velocity of the program star is calculated by measuring the shift in the observed wavelength from the lab wavelength. We have selected a number of clean and unblended lines for this purpose. The radial velocity obtained for the star is $\sim$ 105.3 kms$^{-1}$ (With heliocentric correction).

\subsection{Stellar atmospheric parameters}

Atmospheric parameters of the program star are determined using  the equivalent width measurements of clean and unblended Fe I and Fe II lines (Table\,\ref{tab:2}; The numbers in parentheses in Columns 5 give the derived abundances from the respective line.). The line information are selected from linemake(linemake contains laboratory atomic data (transition probabilities, hyperfine and isotopic substructures) published by the Wisconsin Atomic Physics and the Old Dominion Molecular Physics groups. These lists and accompanying line list assembly software have been developed by C. Sneden and are curated by V. Placco at \url{https://github.com/vmplacco/linemake} \citep{Placco2021}). The lines are identified by overplotting the Arcturus spectra on the spectra of HE 1104$-$0957. A master linelist is then prepared using the line information and the measured equivalent widths. We used MOOG (\cite{Sneden_1973}, updated version 2013) for the analysis under the assumption of Local Thermodynamic Equilibrium (LTE). We have generated the model atmospheres required for the analysis using Kurucz grid of model atmospheres with no convective overshooting \url{http://cfaku5.cfa.hardvard.edu/}.

\begin{table}[t]
\centering
\begin{minipage}{150mm}
\caption{Fe lines used for deriving atmospheric parameters.}
\label{tab:2}
\end{minipage}
\bigskip

\begin{tabular}{ccccc}
\hline   
\textbf{Wavelength} & \textbf{Element} & \textbf{\textit{E}$_{\mathbf{low}}$} & \textbf{log\,\textit{gf}} & \textbf{Equivalent width} \\
\textbf{(\AA)}      &                  & \textbf{(eV)}                    &                           & \textbf{(m\r{A})}         \\
\hline									
4147.670	&	Fe I	&	1.480	&	$-$2.104	&	140.9(4.61)	\\
4153.900	&		&	3.400	&	$-$0.270	&	 70.7(4.44)	\\
4489.740	&		&	0.120	&	$-$3.900	&	172.2(4.73)	\\
4531.150	&		&	1.490	&	$-$2.100	&	123.3(4.18)	\\
4789.650	&		&	3.550	&	$-$0.840	&	 47.9(4.85)	\\
4994.130	&		&	0.920	&	$-$3.080	&	138.8(4.38)	\\
5079.220	&		&	2.200	&	$-$2.100	&	 63.4(4.40)	\\
5194.940	&		&	1.560	&	$-$2.020	&	158.4(4.52)	\\
5195.470	&		&	4.220	& 	$ $0.020	&	 23.9(4.48)	\\
5198.710	&		&	2.220	&	$-$2.135	&	 83.9(4.69)	\\
5247.050	&		&	0.090	&	$-$4.970	&	121.6(4.74)	\\
5250.650	&		&	2.200	&	$-$2.180	&	 81.4(4.66)	\\
5266.560	&		&	3.000	&	$-$0.490	&	125.4(4.66)	\\
6136.610	&		&	2.450	&	$-$1.400	&	101.6(4.40)	\\
6137.690	&		&	2.590	&	$-$1.403	&	 99.6(4.58)	\\
6335.330	&		&	2.200	&	$-$2.180	&	 70.2(4.46)	\\
4223.160	&	Fe II	&	2.580	&	$-$2.000	&	 27.9(4.40)	\\
4508.280	&		&	2.850	&	$-$2.210	&	 17.0(4.64)	\\
\hline
\end{tabular}
\end{table}

Effective temperature and microturbulent velocity are taken to be those values for which there is no trend between the abundances of Fe I and Fe II lines and the corresponding excitation potential, and the reduced equivalent width (W$_{\lambda}$/$\lambda$) respectively. Microturbulent velocity is determined once the effective temperature is fixed. Under these values of temperature and microturbulent velocity, $\log g$ is varied in a number of iterations in such a way that the abundances derived from Fe I and Fe II lines are nearly the same. The detailed procedure can be found in \cite{Purandardas2019b}, and \cite{Purandardas2019a}. The derived atmospheric parameters for our program star are presented in Table\,\ref{tab:3}.

\begin{table}[t]
\centering
\begin{minipage}{156mm}
\caption{Atmospheric parameters of HE 1104$-$0957 derived in this work [1] and literature values from \cite{Anders2022} [2].}
\label{tab:3}
\end{minipage}  
\bigskip

\begin{tabular}{lcccccc}
\hline
\textbf{Star}   &\textbf{\textit{T}$_{\mathbf{eff}}$} & \textbf{log\,\textit{g}} & \textbf{$\zeta$}             & \textbf{[Fe I/H]} & \textbf{[Fe II/H]} & \textbf{Reference}\\
                &\textbf{($\pm$100)}              & \textbf{($\pm$0.2)}      & \textbf{($\pm$0.2)}           &                   &                    & \\
                &\textbf{(K)}                     & \textbf{(cgs)}           & \textbf{(km\,s$^{\mathbf{-1}}$)} &                   &                    & \\
\hline
HE 1104$-$0957  & 3900      & 0.60       & 2.75              & $-$2.95$\pm$0.17     & $-$2.98$\pm$0.17  & [1] \\ 
                & 3983      & 0.69       &  -                & -                    & -                 & [2] \\
\hline
\end{tabular}
\end{table}

\subsection{Abundance analysis}

The procedure adopted for the selection and identification of neutral and ionised lines due to various elements is the same as those employed for Fe lines, as described in Section\,4.  As the lines due to most of the elements in the program star are contaminated by molecular contributions, we have employed spectrum synthesis calculation for the abundance determination. Abundances of 21 elements could be estimated. These include C, N, O, $\alpha$ and Fe-peak elements Mg, Ca, Sc, Ti, V, Cr, Mn, Co, Ni and Zn, and neutron-capture elements Y, Ba, La, Ce, Pr, Nd, Sm, and Eu. The results of the abundance analysis in HE 1104$-$0957 are presented in Table\,\ref{tab:4}. Solar values are taken from \cite{Asplund2009}. The number inside parentheses in column 4 of this table shows the number of lines used for the abundance determination.

\begin{table}
\centering
\begin{minipage}{150mm}
\caption{Elemental abundances in HE 1104$-$0957.}
\label{tab:4}
\end{minipage}
\bigskip

\begin{tabular}{lccccccccc}
\hline                      
& \textbf{Z} & \textbf{solar log$\epsilon$} & \textbf{log$\epsilon$} & \textbf{[X/H]} & \textbf{[X/Fe]} \\
\hline
C (CH, 4315 \AA)     &  6 & 8.43 &    7.30$\pm$0.20(syn)    & $-$1.13 & 1.82    \\
C (C$_{2}$, 5165 \AA) &  6 & 8.43 &    7.32$\pm$0.20(syn)    & $-$1.11 & 1.87    \\
C (C$_{2}$, 5165 \AA) &  6 & 8.43 &    7.39$\pm$0.20(syn)    & $-$1.04 & 1.94    \\
N (CN, 4215 \AA)     &  7 & 7.83 &    7.42$\pm$0.20(syn)    & $-$0.41 & 2.54    \\
O ([OI] 6363.8 \AA)  &  8 & 8.69 &    7.28$\pm$0.20(1, syn) & $-$1.41 & 1.54    \\
Mg I (4571.10 \AA)   & 12 & 7.60 &    4.83$\pm$0.20(1, syn) & $-$2.77 & 0.18    \\
Ca I (5588.7 \AA)    & 20 & 6.34 &    3.80$\pm$0.20(1, syn) & $-$2.54 & 0.41    \\
Sc II (5239.8 \AA)   & 21 & 3.15 &    0.83$\pm$0.20(1, syn) & $-$2.32 & 0.66    \\
Ti I (4555.4 \AA)    & 22 & 4.95 &    2.32$\pm$0.20(1, syn) & $-$2.63 & 0.32    \\
V I (4864.7 \AA)     & 23 & 3.93 &    1.70$\pm$0.20(1, syn) & $-$2.23 & 0.72    \\
Cr I (5345.80 \AA)   & 24 & 5.64 &    3.42$\pm$0.20(1, syn) & $-$2.22 & 0.76    \\
Mn I (4451.5 \AA)    & 25 & 5.43 &    2.97$\pm$0.20(1, syn) & $-$2.46 & 0.49    \\
Fe I                 & 26 & 7.50 &    4.55$\pm$0.17(16)     & $-$2.95 & -       \\
Fe II                & 26 & 7.50 &    4.52$\pm$0.17(2)      & $-$2.98 & -       \\
Co I (5342.7 \AA)    & 27 & 4.99 &    2.60$\pm$0.20(1, syn) & $-$2.39 & 0.56    \\
Ni I (6128.9 \AA)    & 28 & 6.22 &    3.98$\pm$0.20(1, syn) & $-$2.24 & 0.71    \\
Y II (4883.6 \AA)    & 39 & 2.21 &    0.20$\pm$0.20(1, syn) & $-$2.01 & 0.97    \\
Ba II (6141.7 \AA)   & 56 & 2.18 &    0.10$\pm$0.20(1, syn) & $-$2.08 & $<$0.90 \\
La II (4921.7 \AA)   & 57 & 1.10 & $-$1.50$\pm$0.20(1, syn) & $-$2.60 & 0.38    \\
Ce II (4562.3 \AA)   & 58 & 1.58 & $-$0.53$\pm$0.20(1, syn) & $-$2.11 & 0.87    \\
Pr II (5292.6 \AA)   & 59 & 0.72 & $-$1.12$\pm$0.20(1, syn) & $-$1.84 & 1.13    \\
Nd II (5255.5 \AA)   & 60 & 1.42 & $-$0.21$\pm$0.20(1, syn) & $-$1.63 & 1.35    \\
Sm II (4434.3 \AA)   & 62 & 0.96 & $-$0.77$\pm$0.20(1, syn) & $-$1.73 & 1.25    \\ 
Eu II (6645.1 \AA)   & 63 & 0.52 & $-$0.62$\pm$0.20(1, syn) & $-$1.14 & $<$1.83 \\
\hline
\end{tabular}
\end{table}

\subsection{Oxygen, Carbon, Nitrogen}

The abundance of oxygen is derived from the spectrum synthesis calculation of the oxygen forbidden line [OI] 6363.8 \AA. Oxygen is found to be enhanced with [O/Fe] $\sim$ 1.54. At this oxygen abundance, the abundance of carbon is determined from the spectrum synthesis calculation of the CH band at 4315 \AA , C$_{2}$ bands at 5165 \AA, and 5635 \AA. We have used the carbon abundance derived from the CH band throughout the analysis. Spectrum synthesis of CN band at 4215 \AA~is used to derive the nitrogen abundance. Abundances of carbon and nitrogen are also found to be enhanced and the estimated abundances are listed in Table\,\ref{tab:4}.

\subsection{$\alpha$- and Fe-peak elements}

In HE 1104$-$0957, $\alpha$-elements are found to be slightly enhanced with [$\alpha$/Fe] $\sim$ 0.46. Fe-peak elements are also moderately enhanced in the object with a value 0.63 with respect to Fe. The lines used for the synthesis calculations for the abundance determination of these elements are mentioned in Table\,\ref{tab:4}.

\subsection{Neutron-capture elements}

We could estimate the abundances of the neutron-capture elements Y, Ba, La, Ce, Pr, Nd, Sm, and Eu. We could not estimate the abundances of the light s-process elements Sr, and Zr as the lines are heavily blended. In HE 1104$-$0957, r-process elements are found to be more enhanced than the s-process elements. The abundances of these elements as well as the lines used for the abundance determination are listed in Table\,\ref{tab:4}.



\section{Conclusions}

In this paper, we present the abundance analysis results for the object HE 1104$-$0957 based on high-resolution spectra. 
The early classification of this object as a C-R star was based on analysis of spectra obtained with HCT/HFOSC, which is one of the  observing facilities under BINA collaboration. In future, we would also like to use 
other observing facilities under BINA collaboration, such as ARIES-Devasthal Faint Object Spectrograph Camera (ADFOSC) attached to 3.6m Devasthal Optical Telescope (DOT) for low to medium resolution spectroscopy of faint high latitude carbon stars and related objects, in particular for identification and  classification.

Although the low-resolution spectra of the program star show the spectral features as expected for a C-R star, our  analysis shows that the star does not show abundance patterns  characteristic of C-R star. In particular, C-R stars are not known to show an enhanced abundance of r-process elements such as Eu. From our analysis, we find that the object HE 1104$-$0957 is an extremely metal-poor giant with a metallicity [Fe/H] $\sim$ {\bf $-$}2.98 with very peculiar elemental abundance patterns. Although, carbon is found to be enhanced in HE 1104$-$0957, it does not fall in any of the sub-groups of CEMP stars based on the characteristic elemental abundance ratios used for the classification of various CEMP sub-groups. Other than carbon, nitrogen and oxygen are also found to be enhanced in the star. The abundance analysis results show that $\alpha-$, and Fe-peak elements are slightly enhanced. Neutron-capture elements are also found to be enhanced in the star. Among the neutron-capture elements, r-process elements are more enhanced than the s-process elements.

HE 1104$-$0957 shows enhancement of Eu with [Eu/Fe] $\leq$ 1.83. Previous studies have shown that metal-poor stars with this level of r-process enhancement are only rarely found in the halo. Based on the observed abundance patterns, we expect that the origin of this object may be traced in neighbouring dwarf galaxies, and it needs to be further investigated. Hence, a comprehensive study  is required to draw a robust conclusion in this regard. Such a study is in progress, and the results will be presented in a future paper.

\begin{acknowledgments}
MP and AG would like to thank  the organizers for their kind hospitality during the third BINA workshop. MP is grateful for financial support from the Indian Institute of Astrophysics, Bangalore. AG  acknowledges  the support received from the Belgo-Indian Network for Astronomy \&
Astrophysics  project BINA - 2  (DST/INT/Belg/P-02 (India) and BL/11/IN07 (Belgium)).
\end{acknowledgments}

\begin{furtherinformation}

\begin{orcids}
\orcid{0000-0001-5047-5950}{Meenakshi}{Purandardas}
\orcid{0000-0002-8841-8641}{Aruna}{Goswami}
\end{orcids}

\begin{authorcontributions}
This work is part of a collective effort with contributions from all the co-authors.
\end{authorcontributions}

\begin{conflictsofinterest}
The authors declare no conflict of interest.
\end{conflictsofinterest}

\end{furtherinformation}

\bibliographystyle{bullsrsl-en}

\bibliography{S05-P03_PurandardasM}



\end{document}